\documentclass[12pt]{article}             
\begin{document}
\begin{center}
{\bf No signaling and strong subadditivity condition for tomographic q-entropy of single qudit states}\\
V. N. Chernega$^{1}$, O.~V.~Man'ko$^{1,2}$\\
1 - P.N.~Lebedev Physical Institute, Leninskii Prospect, 53, Moscow 119991, Russia \\
2 - Bauman Moscow State Technical University, The 2nd Baumanskaya Str. 5, Moscow\\
      Email: omanko@sci.lebedev.ru
\end{center}




\vspace{10pt}

\begin{abstract}
The single qudit state tomograms are shown to have the no signaling property. The known and new entropic and information inequalities for Shannon, von Neumann and $q$-entropies of the composite and noncomposite systems characterizing correlations in these systems are discussed. The spin tomographic probability distributions determining the single qudit states are demonstrated to satisfy the strong subadditivity condition for Tsallis $q$-entropy. Examples of the new entropic inequalities for $q$-entropy are considered for qudits with $j=5/2$, $j=7/2$.
\end{abstract}
\vspace{2pc}
\vspace{0.4cm}

\section{Introduction}
The quantum correlations like entanglement phenomenon associated with properties of multipartite system states provide also specific entropic and information inequalities for such systems. In our study we consider entropic properties of systems without subsystems. The aim of the work is to review and obtain some new $q$-entropic inequalities for single qudit state tomogram including strong subadditivity condition for noncomposite quantum system and to demonstrate the no signaling property for such system state. The subadditivity and strong subadditivity conditions for Shannon \cite{Shanon} and von Neumann \cite{vonNeuman} entropies are known for bipartite and three-partite system states, respectively, both for classical and quantum ones. Recently, the approach was suggested \cite{maps,OlgaVova1,RitaPS2014,newineq,subadditivity,Fedorov} to extend the known for composite systems entropic and information relations like equalities and inequalities to the systems without subsystems. Among these relations there are inequalities for Tsallis entropy of the bipartite systems and other inequalities for von Neumann entropy proved and studied in \cite{RitaPS2014,Furuci,Furuci1,Petz-Vi,fromRusk,LiebRuskai,NielsonCH,PetzNielson,CorlenLieb2007,20pr,20pra,20prb,20prc,2014,24a,24b,24c,24d,24e,24n,Kh,TurinVova,TurinMA,OlgaVova}. The idea of this approach is to use map of integers $N$ on pairs of integers $(j k)$ or triples of integers $(j k l)$, etc. This map of integers provides the possibility to consider vectors $\vec p$ with components $p_s$ as the matrix ${\cal P}$ with matrix elements ${\cal P}_{j k}$, i.e. we map the indice $s$ onto combined index $s\leftrightarrow (j k)$ or interpret the index $s$ as the function $s\equiv s(j k)$. Analogously the matrices $\rho_{s s'}$ can be considered as matrices $\rho_{j k, j' k'}$ since due to the described map the integers $s$ and $s'$ are functions of the two variables $s(j k)$ and $s'(j' k')$. Also the vector with components $p_s$ can be considered as the table of numbers $p_{j k l}$ due to invertable map $s\leftrightarrow (j k l)$. In this case the matrix $\rho_{s s'}$ can be considered as matrix $\rho_{j k l, j'k'l'}$. After such map one can apply the analog of partial tracing procedure used for probability vectors or joint probability distributions and density matrices of composite systems to get the reduced matrices. For specific maps such tool was called portrait method \cite{Vovf,Lupo} to be applied for studying entanglement of multiqudit states. The joint probability $P(a b|x y )$ of two random variables $a,b$ depending on two extra parameters $x$ and $y$ has the no signaling property which means that $\sum_b P(a b|x y )$ does not depend on parameter $y$ and $\sum_a P(a b|x y )$ does not depend on parameter $x$. The tomogram \cite{DodPLA,OlgaJETP} of two qudit states has this property \cite{maps}.
We will discuss this no signaling phenomenon also for single qudit state tomograms. The strong subadditivity condition was proved for von Neumann (and Shannon) entropy of three partite system \cite{LiebRuskai}. It is inequality for entropies of the system state and entropies of its three different subsystems. This inequality is not valid \cite{Petz-Vi} for Tsallis quantum entropy \cite{Tsallis}. The new result which we present in the work is to show that the $q$-entropies associated with tomograms of quantum states both for three qudit system and for single qudit system obey the strong subadditivity condition. \\

\noindent The paper is organized as follows.\\

\noindent In the second section, we study no signaling property of two qudit state tomograms.
In the third section, we discuss the no signaling properties of single qudit state tomograms. In the fourth section we consider an example of qudit with $j=5/2$ and study its no signaling properties. In the
fifth section, we discuss the q-deformed strong subadditivity condition in the case of
single qudit state. In conclusion, we list our main results. In Appendix we present new formulas for arbitrary nonnegative Hermitian matrices.

\section{No signaling property of the two qudit state tomograms}
The tomogram $w(m_1,m_2|u)$ of two qudit state with spins $j_1$ and $j_2$ and with density matrix $\rho(1,2)_{m_1m_2,{m'}_1{m_2}'}$ where spin projections $m_1({m')}_1$, $m_2({m')}_2$ take semiinteger values $-j_1,-j_1+1,\ldots j_1-1,j_1$ and $-j_2,-j_2+1,\ldots j_2-1,j_2$, respectively is the fair probability distribution defined as set of diagonal matrix elements of the matrix $u\rho(1,2)u^\dag$, i.e.
\begin{equation}\label{eq.1}
w(m_1,m_2|u)=(u\rho(1,2)u^\dag)_{m_1m_2,m_1m_2}.
\end{equation}
Here $u$ is unitary matrix describing the global unitary transform in the Hilbert space ${\cal H}=H_1\times H_2$ and the Hilbert spaces $H_1$ and $H_2$ are the spaces of states for first and second qudits, respectively. The dimension of the space $H$ equals $N=n m$ where dimensions of spaces $H_1$ and $H_2$ are equal to integers $n$ and $m$. One has relations $n=2j_1+1$, $m=2j_2+1$. If the matrix $u$ equals to tensor product of unitary $n\times n$-matrix $u_1$ and unitary $m\times m$-matrix $u_2$, i.e. $u=u_1\otimes u_2$ the tomogram of the quantum state reads
\begin{equation}\label{eq.2}
w(m_1,m_2|u_1,u_2)=(u_1\otimes u_2\rho(1,2)u_1^\dag\otimes u_2^\dag)_{m_1m_2,m_1m_2}.
\end{equation}
If the matrices $u_1$ and $u_2$ are matrices of irreducible representations of the group SU(2) corresponding to spins $j_1$ and $j_2$, respectively the tomogram of the quantum state of the qudits is spin-tomogram
\begin{equation}\label{eq.3}
w(m_1,m_2|\vec n_1,\vec n_2)=(u_1(\vec n_1)\otimes u_2(\vec n_2)\rho(1,2)u_1^\dag(\vec n_1)\otimes u_2^\dag(\vec n_2))_{m_1m_2,m_1m_2}.
\end{equation}
Here $\vec n_1$ and $\vec n_2$ are unit vectors depending on angles $\phi_1,\theta_1$ and $\phi_2,\theta_2$ perpendicular to the Poincare sphere. The  physical meaning of the tomogram (\ref{eq.3}) is the following. It equals to joint probability distribution to get the spin projections $m_1$ and $m_2$ on the quantization axes determined by vectors $\vec n_1$ and $\vec n_2$. From the physical meaning of the spin tomogram follows the relation for the marginal probability distribution
$w(m_1|u)=\sum_{m_2=-j_2}^{j_2}w(m_1,m_2|u)$
and the relation is
\begin{equation}\label{eq.5}
w_1(m_1|u_1\otimes u_2)\equiv w_1(m_1|u_1).
\end{equation}
For spin tomogram (\ref{eq.3}) one has
\begin{equation}\label{eq.6}
w_1(m_1|\vec n_1,\vec n_2)=\sum_{m_2=-j_2}^{j_2}w(m_1,m_2|\vec n_1,\vec n_2)\equiv w_1(m|\vec n_1).
\end{equation}
It means that such parameters of second subsystem (second qudit) like $u_2$ or $\vec n_2$ do not determine the tomographic probability distribution of the first spin (first qudit). This property is called no signaling property. On the other hand, this obvious from the point of view of physical properties of two random observables, feature is a numerical relation for matrix elements of the matrix $u\rho(1,2)u^\dag$. This observation provides the possibility to find analog of the no signaling property for other systems, e.g. for single qudit state with spin $j$. We follow the approach \cite{Furuci,Furuci1,Petz-Vi,fromRusk,LiebRuskai,NielsonCH,PetzNielson,CorlenLieb2007,20pr,2014,RitaPS2014,TurinVova,TurinMA,OlgaVova} to use the bijective maps of integers or semiintegers to get the no signaling property for system without subsystems.

\section{No signaling for single qudit state tomogram}.
Given density $N\times N$-matrix $\rho$ of single qudit state where $N=n m=2j+1$. Let us use the map |$-j\leftrightarrow1,$ $-j+1\leftrightarrow2,\,\ldots,\,j-1\leftrightarrow N-1\,j\leftrightarrow N$. The matrix elements of density matrix of the single qudit state in this case are $\rho_{\alpha\beta}$, where $\alpha,\beta=1,2\,\ldots,\,N.$ The matrix $\rho$ can be presented in block form
\begin{equation}\label{eq.7}
\rho=\left(\begin{array}{cccc}
\rho_{11}&\rho_{12}&\dots&\rho_{1n}\\
\rho_{21}&\rho_{22}&\dots&\rho_{2n}\\
\dots&\dots&\dots&\dots\\
\rho_{n1}&\rho_{n2}&\dots&\rho_{nn}
\end{array}\right).
\end{equation}
Here the blocks $\rho_{j k}$ with $j,k=1,2,\ldots,n$ are $m\times m$-matrices. One can construct two matrices $\rho_1$ and $\rho_2$ using the blocks $\rho_{j k}$. The $n\times n$ - matrix $\rho_1$ has the matrix elements $(\rho_1)_{j k}=\mbox{Tr}\rho_{j k}$. The $m\times m$-matrix $\rho_2=\sum_{k=1}^n\rho_{k k}.$ In case of two-qudit state with density matrix $\rho\equiv\rho(1,2)$, considered in previous section, the matrix $\rho_1$ coincides with density matrix of first qudit state and the matrix $\rho_2$ coincides with the density matrix of second qudit state. In this case the tomogram $w_1(m_1|u_1)$ is determined by the diagonal matrix elements of the matrix $(u_1\rho_1u_1^\dag)_{m_1m_1}$ and the tomogram $w_2(m_2|u_2)=(u_2\rho_2u_2^\dag)_{m_2m_2}$.  On the other hand the tomographic probability  distribution which determines the density matrix $\rho$ of single qudit state equals to diagonal matrix element of the $N\times N$-matrix $u\rho u^\dag$. Since available numerical relations for the matrix elements of the matrices do not depend o n any interpretation of the matrices we can write these relations for the Hermitian nonnegative $N\times N$ matrix $\rho$ with $\mbox{Tr}\rho=1$ and unitary matrices $u,\,u_1$ and $u_2$. The relations can be presented in the form of relations for probability vector $\vec w_\rho(u)$ with $N$ components equal to diagonal matrix elements of the matrix $u\rho u^\dag$. The probability vector reads \cite{probability vector}
\begin{equation}\label{eq.8}
\vec w_\rho(u)=|u u_0|^2\vec\rho.
\end{equation}
Here $\vec\rho$ is $N$-vector with components equal to eigenvalues of the matrix $\rho$. The unitary $N\times N$ matrix $u_0$ has the columns which are corresponding eigenvectors of the matrix $\rho$. The notation for $N\times  N$-matrix $|a|^2$ means that the matrix elements of the matrix ${|a|^2}_{\alpha\beta}=|a_{\alpha\beta}|^2$. Let us introduce stohastic $N\times N$-matrix $M^{(1)}$ given in the block form analogous to (\ref{eq.7}), i.e.
\begin{equation}\label{eq.9}
M^{(1)}=\left(\begin{array}{cccc}
M_{11}^{(1)}&M_{12}^{(1)}&\dots&M_{1n}^{(1)}\\
M_{21}^{(1)}&M_{22}^{(1)}&\dots&M_{2n}^{(1)}\\
\dots&\dots&\dots&\dots\\
M_{n1}^{(1)}&M_{n2}^{(1)}&\dots&M_{nn}^{(1)}
\end{array}\right).
\end{equation}
The blocks $M^{(1)}_{j k}$ for $j\geq 2$ are zero $m\times m$-matrices. The blocks $M_{1 k}^{(1)}$ have all nonzero matrix elements equal to one in $k$th row. The probability $N$-vector
\begin{equation}\label{eq.10}
\vec w_1(u)=M^{(1)}|u u_0|^2\vec\rho
\end{equation}
has the property
\begin{equation}\label{eq.11}
\vec w_1(u_1\otimes u_2)=M^{(1)}|(u_1\otimes 1) u_0|^2\vec\rho.
\end{equation}
In (\ref{eq.11}) the unitary matrix $u_2=1$ is $m\times m$-matrix. This probability vector (\ref{eq.11}) has only $n$ first nonzero components. The $n$-vector with these components is the tomographic probability distribution which determines the $n\times n$-matrix $\rho_1$. The components of the vector are equal to diagonal elements of the $n\times n$-matrix $u_1\rho_1u_1^\dag$.

Analogously, we introduce the stohastic $N\times N$-matrix $M^{(2)}$ of the block form with blocks $M^{(2)}_{j k}$ where the only nonzero blocks are blocks $M_{1 k}^{(2)}$ which are equal to unity matrices. The probability $N$-vector determined by the matrix $M^{(2)}$ reads
\begin{equation}\label{eq.12}
\vec w_2(u)=M^{(2)}|u u_0|^2\vec\rho.
\end{equation}
It has first $m$ nonzero components. The property of this vector (\ref{eq.12}) analogous to (\ref{eq.11}) is
\begin{equation}\label{eq.13}
\vec w_2(u_1\otimes u_2)=M^{(2)}|(1\otimes u_2) u_0|^2\vec\rho.
\end{equation}
In (\ref{eq.13}) the unitary matrix $u_1=1$ is $n\times n$-matrix. The independence of tomographic probability vectors $\vec w_1(u_1\otimes u_2)$ and $\vec w_2(u_1\otimes u_2)$ on the unitary transforms $u_2$ and $u_1$, respectively, reflects no signaling property. For the bipartite system the vectors are just tomographic probability distributions of the subsystem states.

Thus we conclude that for any single qudit quantum state with density $N\times N$ matrix $\rho$, if $N=n m$, there exist properties of probability distribution associated with probability vector (\ref{eq.8}) which are analogous to no signaling properties of the tomograms of bipartite system states. We illustrate the presented results on examples of qudit with $j=5/2$ and analogous bipartite qubit-qutrit composite system.

\section{Example of no signaling for qudit $j=5/2$}
Let $6\times6$-matrix $\rho$ be the density matrix of a qudit a state for $j=5/2$ and it is presented in block form (\ref{eq.7}), i.e. for $n=2$, $m=3$ one has
\begin{equation}\label{eq.14}
\rho=\left(\begin{array}{cc}
\rho_{11}&\rho_{12}\\
\rho_{21}&\rho_{22}
\end{array}\right).
\end{equation}
where the blocks $\rho_{j k}$, ($j,\,k=1,2$) are $3\times3$-matrices. In this case the stohastic $6\times6$-matrices  $M^{(1)}$ and $M^{(2)}$ read
\begin{equation}\label{eq.15}
M^{(1)}=\left(\begin{array}{cccccc}
1&1&1&0&0&0\\
0&0&0&1&1&1\\
0&0&0&0&0&0\\
0&0&0&0&0&0\\
0&0&0&0&0&0\\
0&0&0&0&0&0
\end{array}\right),\quad M^{(2)}=\left(\begin{array}{cccccc}
1&0&0&1&0&0\\
0&1&0&0&1&0\\
0&0&1&0&0&1\\
0&0&0&0&0&0\\
0&0&0&0&0&0\\
0&0&0&0&0&0
\end{array}\right).
\end{equation}
The $2\times2$-matrix $\rho_1$ and $3\times 3$-matrix $\rho_2$ are
\begin{equation}\label{eq.16}
\rho_1=\left(\begin{array}{cc}
\mbox{Tr}\rho_{11}&\mbox{Tr}\rho_{12}\\
\mbox{Tr}\rho_{21}&\mbox{Tr}\rho_{22}
\end{array}\right), \quad \rho_2=\rho_{11}+\rho_{22}.
\end{equation}
The $6$-vector $\vec w(u)$ has the components \[(w(+5/2|u),\,w(+3/2|u),\,\,w(+1/2|u),\,\,w(-1/2|u),\,w(-3/2|u),\,w(-5/2|u))\]
which are diagonal elements of matrix $u\rho u^\dag$. We used natural notation $w(\bar m|u)$ where $\bar m$ is spin projection value in reference frame rotated by means of unitary $6\times6$-matrix in the Hilbert space $H$ of the qudit states. It means that we use map of indices $1\leftrightarrow+5/2$, $2\leftrightarrow+3/2$, $3\leftrightarrow+1/2$, $4\leftrightarrow-1/2$, $5\leftrightarrow-3/2$, $6\leftrightarrow-5/2$ to label the matrix elements of the density matrix $\rho_{\alpha\beta}$, $(\alpha,\beta=1,2,\ldots,6)$. The tomogram of the matrix $\rho$ which is $6$-vector $\vec w(u)$ gives two $6$-vectors $\vec w_1(u)=M^{(1)}\vec w(u)$ and $\vec w_2(u)=M^{(2)}\vec w(u)$. The $6$-vector $\vec w_1(u)$ has two nonzero first components, the $6$-vector $\vec w_2(u)$ has three nonzero first components. The nonzero components provide $2$-vector and $3$-vector which are analogs of vectors given by tomographic probability distributions associated with matrices $\rho_1$ and $\rho_2$, respectively. By construction these probability vectors correspond to marginal probability distributions associated with artificial joint probability distribution given by $6$-vector $\vec w(u)$. The $6$-vectors $\vec w_1(u)$ and $\vec w_2(u)$ have the no signaling properties, i.e. if $u=u_1\otimes u_2$ one has
\begin{equation}\label{eq.17}
\vec w_1(u_1\otimes u_2)\equiv\vec w_1(u_1\otimes1),\quad\vec w_2(u_1\otimes u_2)\equiv\vec w_2(1\otimes u_2).
\end{equation}
These properties can be proved by direct checking. For qubit-qutrit bipartite system with density matrix $\rho$ (\ref{eq.14}) the matrix $\rho_1$ is the density matrix of the qubit state and the matrix $\rho_2$ is the density matrix of qutrit state. These matrices are connected with the matrix $\rho$ by partial tracing procedure. Nonzero components of the $6$-vectors $\vec w_1(u_1\otimes1)$ and $\vec w_2(u_1\otimes u_2)$ provide in this case the tomographic probability vectors with components obtained as diagonal elements of the matrix $u_1\rho_1u_1^\dag$ and $u_2\rho_2u_2^\dag$, respectively.

\section{Strong subadditivity condition for $q$-entropy of single qudit state}
For composite three-partite system with the diagonal density matrix  $\rho(1,2,3)$ it is known \cite{Furuci,Furuci1} that Tsallis $q$-entropy satisfies strong subadditivity condition. It means that for $q\geq 1$
\begin{equation}\label{eq.1a}
S_q(\rho(1,2,3))+S_q(\rho(2))\leq S_q(\rho(1,2))+S_q(\rho(2,3)).
\end{equation}
Here $\rho(1,2)$, $\rho(2,3)$ and $\rho(2)$ are diagonal density matrices determined by partial tracing procedure
\begin{equation}\label{eq.2a}
\rho(1,2)=\mbox{Tr}_3\rho(1,2,3),\quad \rho(2,3)=\mbox{Tr}_1\rho(1,2,3),\quad \rho(2))=\mbox{Tr}_3\rho(2,3).
\end{equation}
In fact, the diagonal matrix elements of the matrix $\rho(1,2,3)$ provide classical joint probability distribution of three random variables. For any density matrix the $q$-entropy is defined as
\begin{equation}\label{eq.3a}
S_q(\rho)=-\mbox{Tr}\rho^q\frac{\rho^{1-q}-1}{1-q},
\end{equation}
where $q\geq1$. We extend the inequality (\ref{eq.1a}) which is valid for composite system to the case of noncomposite system. We will do this for tomographic probability distribution (qudit tomogram) $w(m|u)$ or probability vector $\vec w(u)$, given by (\ref{eq.8}). Let $2j+1=N=n_1n_2n_3$, where $n_k$ are integers. Let us use the bijective map of integers onto spin projections $1,\,2,\,\ldots,\,N\,\leftrightarrow -j,\,-j+1\,\ldots,j-1,\,j.$ Then we introduce the map of the integers onto triples of integers, i.e. $s\leftrightarrow s(i k l)$ where $s=1,2,\ldots,N$, $i=1,2,\ldots,n_1$, $k=1,2,\ldots, n_2$, $l=1,2,\ldots,n_3$. It means that we introduce function of three variables $s(i k l)$.

Thus the probability vector $\vec w(u)$ with $N$ components $w_s(u)$ can be considered as the probability vector with components $\vec w(u)_{s(i k l )}$. Then one can construct three probability $N$-vectors $\vec w_{12}(u)$, $\vec w_{23}(u)$, $\vec w_{2}(u)$ with nonzero components
\begin{eqnarray}
&&(\vec w_{12}(u))_{s(i k)}=\sum_{l=1}^{n_3}(\vec w(u))_{s(i k l )},\quad(\vec w_{23}(u))_{s( kl)}=\sum_{i=1}^{n_1}(\vec w(u))_{s(i k l )}\nonumber\\
&&(\vec w_{2}(u))_{s(k)}=\sum_{i=1}^{n_1}\sum_{l=1}^{n_3}(\vec w(u))_{s(i k l )}.\label{eq.4a}
\end{eqnarray}
Other components of the probability vectors equal to zero. The strong subadditivity condition for the quantum tomogram $w(m|u)$ written in terms of the probability vector components reads
\begin{eqnarray}
&&-\frac{1}{1-q}\sum_{i=1}^{n_1}\sum_{k=1}^{n_2}\sum_{l=1}^{n_3}\left(\vec w(u)\right)^q_{s(i k l)}\left[(\vec w(u))^{1-q}_{s(i k l )}-1\right]
-\frac{1}{1-q}\sum_{k=1}^{n_2}\left(\vec w_2(u)\right)^q_{s( k )}\left[(\vec w(u))^{1-q}_{s( k  )}-1\right]\leq\nonumber\\
&&
-\frac{1}{1-q}\sum_{k=1}^{n_2}\sum_{l=1}^{n_3}\left(\vec w_{23}(u)\right)^q_{s( k l)}\left[(\vec w_{23}(u))^{1-q}_{s( k l )}-1\right]
-\frac{1}{1-q}\sum_{i=1}^{n_1}\sum_{k=1}^{n_2}\left(\vec w_{12}(u)\right)^q_{s(i k )}\left[(\vec w_{12}(u))^{1-q}_{s(i k )}-1\right].\nonumber\\
&&\label{eq.5a}
\end{eqnarray}
The tomogram of the single qudit state satisfies also the subadditivity condition for $q$-entropy. If we use notation $n_1=n,\,n_2 n_3=m,\,N=n m$, the condition reads
\begin{eqnarray}
&&-\frac{1}{1-q}\sum_{i=1}^{n}\sum_{k=1}^{m}\left(\vec w(u)\right)^q_{s(i k)}\left[(\vec w(u))^{1-q}_{s(i k)}-1\right]\leq
-\frac{1}{1-q}\sum_{i=1}^{n}\left(\vec \Omega_1(u)\right)^q_{s(i )}\left[(\vec \Omega_1(u))^{1-q}_{s(i)}-1\right]\nonumber\\
&&-\frac{1}{1-q}\sum_{k=1}^{m}\left(\vec \Omega_2(u)\right)^q_{s(k )}\left[(\vec \Omega_2(u))^{1-q}_{s(k)}-1\right].
\label{eq.6a}
\end{eqnarray}
Here we introduce map of integers $s=1,2,\ldots,N$ onto pairs of integers $s\leftrightarrow (i k)$ where $i=1,2,\ldots n$, $k=1,2,\ldots m$. It means that we introduce the function $s(i k)$. The nonzero components of probability N-vectors $\vec \Omega_1(u)$ and $\vec \Omega_2(u)$ are defined as
\begin{equation}\label{eq.7a}
\left(\vec \Omega_1(u)\right)_{s(i )}=\sum_{k=1}^{m}\left(\vec w(u)\right)_{s(i k )},\quad
\left(\vec \Omega_2(u)\right)_{s(k )}=\sum_{i=1}^{n}\left(\vec w(u)\right)_{s(i k )}.
\end{equation}
The subadditivity condition (\ref{eq.6a}) corresponds to the known for bipartite systems $q$-entropy subadditivity condition for matrix $\rho$ and matrices $\rho_1$ and $\rho_2$
\begin{equation}\label{eq.8a}
-\mbox{Tr}\,\rho^q\,\frac{\rho^{1-q}-1}{1-q}\leq
-\mbox{Tr}\,\rho_1^q\,\frac{\rho_1^{1-q}-1}{1-q}-\mbox{Tr}\,\rho_2^q\,\frac{\rho_2^{1-q}-1}{1-q}.
\end{equation}
The matrix $\rho$ is density matrix of the single qudit state. The matrices $\rho_1$ and $\rho_2$ are obtained by means of analog of partial tracing procedure used for composite quantum systems. Also one can get the analogous inequality using change of notation $n_1n_2=n,\, n_3=m$. If $N\neq n_1n_2n_3 $ one can introduce the matrix
\[\bar\rho=\left(\begin{array}{cc}
\rho&0\\0&0\end{array}\right),\]
where $\bar N=N+k=n_1n_2n_3$ and choosing the corresponding integer $k$. The tomogram $\bar w(m|ué
)$ will satisfy in this case inequlity (\ref{eq.6a}) and (\ref{eq.8a}) with obvious substitutions $w\rightarrow\bar w$ in these formulas.

\section{Example of strong subadditivity condition for qudit with $j=7/2$}
Let us consider the density metrix $\rho_{m m'}$ where \[m, m'=-7/2,-5/2,-3/2,-1/2,+1/2,+3/2,+5/2,+7/2.\] The tomogram $w(m|u)$ of the quantum state with this matrix where $u$ is unitary $8\times 8$-matrix is described by the probability $8$-vector (\ref{eq.8}) $\vec w(u)$ with components $w(m|u)$. Let us introduce three stochastic matrices $M^{(12)}$, $M^{(23)}$ and matrix $M^{(2)}$ of the form
\begin{equation}\label{eq.9a}
M^{(12)}=\left(\begin{array}{cccccccc}
1&1&1&1&0&0&0&0\\
0&0&0&0&1&1&1&1\\
0&0&0&0&0&0&0&0\\
0&0&0&0&0&0&0&0\\
0&0&0&0&0&0&0&0\\
0&0&0&0&0&0&0&0\\
0&0&0&0&0&0&0&0\\
0&0&0&0&0&0&0&0\\
\end{array}\right),\quad M^{(23)}=\left(\begin{array}{cccc}
1_4&1_4\\
0_4&0_4\\
\end{array}\right)
\end{equation}
and
\begin{equation}\label{eq.10a}
M^{(2)}=\left(\begin{array}{cccccccc}
1&1&0&0&1&1&0&0\\
0&0&1&1&0&0&1&1\\
0&0&0&0&0&0&0&0\\
0&0&0&0&0&0&0&0\\
0&0&0&0&0&0&0&0\\
0&0&0&0&0&0&0&0\\
0&0&0&0&0&0&0&0\\
0&0&0&0&0&0&0&0\\
\end{array}\right).
\end{equation}
Here $0_4$ is zero $4\times4$-matrix and $1_4$ is identity $4\times4$-matrix. New inequality which is the strong subadditivity condition for the tomographic probability vector $\vec w(u)$ of the single qudit state reads
\begin{eqnarray}
&&-\frac{1}{1-q}\sum_{m=-7/2}^{7/2}\left[\left(\vec w(u)\right)_m\right]^q\left[\left[\left(\vec w(u)\right)_m\right]^{1-q}-1\right]\nonumber\\
&&
-\frac{1}{1-q}\sum_{m=-7/2}^{7/2}\left[\left(M^{(2)}\vec w(u)\right)_m\right]^q
\left[\left[\left(M^{(2)}\vec w(u)\right)_m\right]^{1-q}-1\right]\leq\nonumber\\
&&-\frac{1}{1-q}\sum_{m=-7/2}^{7/2}\left[\left(M^{(12)}\vec w(u)\right)_m\right]^q
\left[\left[\left(M^{(12)}\vec w(u)\right)_m\right]^{1-q}-1\right]\nonumber\\
&&
-\frac{1}{1-q}\sum_{m=-7/2}^{7/2}\left[\left(M^{(23)}\vec w(u)\right)_m\right]^q
\left[\left[\left(M^{(23)}\vec w(u)\right)_m\right]^{1-q}-1\right].
\label{eq.11a}
\end{eqnarray}
From (\ref{eq.11a}) follows new entropic inequalitity
\begin{eqnarray}
&&-\frac{1}{1-q}\mbox{Tr}\rho^q\left(\rho^{1-q}-1\right)-\frac{1}{1-q}\mbox{Tr}\rho_2^q\left(\rho_2^{1-q}-1\right)\leq \nonumber\\ &&-\frac{1}{1-q}\sum_{m=-7/2}^{7/2}\left[\left(M^{(12)}\vec w(u)\right)_m\right]^q
\left[\left[\left(M^{(12)}\vec w(u)\right)_m\right]^{1-q}-1\right]\nonumber\\
&&-\frac{1}{1-q}\sum_{m=-7/2}^{7/2}\left[\left(M^{(23)}\vec w(u)\right)_m\right]^q
\left[\left[\left(M^{(23)}\vec w(u)\right)_m\right]^{1-q}-1\right].
\label{eq.12a}
\end{eqnarray}
In (\ref{eq.12a}) the $2\times2$-matrix $\rho_2$ has matrix elements
\[(\rho_2)_{11}=\rho_{7/2\,7/2}+\rho_{5/2\,5/2}+\rho_{-1/2\,-1/2}+\rho_{-3/2\,-3/2},\quad(\rho_2)_{22}=1-(\rho_2)_{11},\] \[(\rho_2)_{12}=(\rho_2)_{21}^\ast=\rho_{-7/2\,-3/2}+\rho_{-5/2\,-1/2}+\rho_{1/2\,5/2}+\rho_{3/2\,7/2}\]
In right-hand side of the equation (\ref{eq.12a}) the unitary matrix $u$ is arbitrary unitary matrix of the product form $u=u_1\times u_2\times u_3$, where the local transform matrices correspond to integers $n_1,\,n_2\,,n_3$. For $q\rightarrow1$ one has the entropic inequality for the von Neumann entropy associated with the tomogram of the single qudit state.

\section{Conclusion}
To resume we list the main results of our work. We show that the tomographic probability distribution determining the single qudit state has no signaling property which was known for joint tomographic probability distribution. We considered the no signaling property for the tomogram of the qudit (spin) state with $j=5/2$. We found new entropic inequality for single qudit state and this inequality provides the relation for Tsallis $q$-entropy associated with qudit-state tomogram and probability vectors obtained from this tomographic probability vector by action of stochastic matrices. In the case of three-partite system the inequality coincides with strong subadditivity condition for $q$-entropies associated with tomographic probability vectors of the system and its three subsystems. As partial case of the obtained inequality we derived the inequality for the sum of Tsallis quantum entropy of the single qudit state and the analog of one of the subsystems of the system. This new inequality in the limit $q\rightarrow1$ is compatible with analog of the strong subadditivity condition available for Shannon entropies of three-partite system. The new inequalities obtained in this work are given by (\ref{eq.11a})-(\ref{A1}).

\subsection*{Acknowledgements}
We are grateful to Margarita Aleksandrovna and Vladimir Ivanovich Man'ko, our parents and grandparents for their supervising during whole our life starting from the very beginning.

\section{Appendix}
We present the new inequality which is valid for arbitrary $N\times N$ - matrix $\rho$, where $N=nm$ and $\rho^\dag=\rho$, $\mbox{Tr}\rho=1$. The matrix $\rho$ is nonnegative and we consider also arbitrary unitary $N\times N$- matrix $u$. Then for $q\geq1$ one has inequality
\begin{eqnarray}\label{A1}
&&1+\sum_{\alpha=1}^N\left[\left(u\rho u^\dag\right)_{\alpha\alpha}\right]^q\leq\sum_{\beta=1}^N\left\{\left[\sum_{\alpha=1}^N M_{\beta\alpha}^{(1)}\left(u\rho u^\dag\right)_{\alpha\alpha}\right]^q+\left[\sum_{\alpha=1}^N M_{\beta\alpha}^{(2)}\left(u\rho u^\dag\right)_{\alpha\alpha}\right]^q\right\}.\nonumber\\&&\label{A1}
\end{eqnarray}
Here the stohastic $N\times N$-matrices $M^{(1)}$ and $M^{(2)}$ are given in the block form (\ref{eq.9}) with $m\times m$-blocks $M_{j k}^{(1)}$ and $M_{j k}^{(2)}$, $j,k=1,2,\ldots,n$. All the $m\times m$-blocks
$M_{j k}^{(1)}$ and $M_{j k}^{(2)}$ for $j\geq 2$ are zero matrices. The blocks $M_{j k}^{(1)}$ have nonzero matrix elements equal to 1 only in the $k$th rows for $j=1$. All the blocks $M_{j k}^{(2)}$ are either zero or unity $m\times m$-matrices. For example, the $6\times6$-matrices $\rho$ and $u$ satisfy the inequality (\ref{A1}) for $q\geq1$ where the matrices $M^{(1)}$ and $M^{(2)}$ are given by Eq.(\ref{eq.15}). The written matrix inequality gives the subadditivity condition for the system state tomogram in case where the matrix $\rho$ is a density matrix of a bipartite system, e.g. of two qudit system with $2j_1+1=n$, $2j_2+1=m$.


\end{document}